\documentclass[journal]{IEEEtran}
\usepackage{cite}

\usepackage{enumerate}
\usepackage{graphicx} 
\usepackage{multirow} 
\usepackage{amsmath,amsfonts,amssymb}

\usepackage{epstopdf} 

\usepackage{url}
\usepackage{xcolor}
\usepackage{threeparttable}
\usepackage[perpage,symbol]{footmisc}
\setfnsymbol{wiley}

\usepackage{CJK}
\ifCLASSINFOpdf
\else
\fi
\begin{document}

%
\title{HCIC: Hardware-assisted Control-flow Integrity Checking}
%
%
%
%
\author{Jiliang~Zhang,~\IEEEmembership{Member,~IEEE,}~Binhang Qi,~Zheng Qin,~Gang~Qu

\thanks{This work is supported by the National Natural Science Foundation of China (Grant NOs. 61874042, 61602107), the National Natural Science Foundation of Hunan Province, China (Grant No. 618JJ3072), the Hu-Xiang Youth Talent Program, the 2017 CCF-IFAA RESEARCH FUND, and the Fundamental Research Funds for the Central Universities.}


\thanks{\textcolor[rgb]{1.00,0.00,0.00}{This paper has been accepted by the IEEE Internet of Things Journal, Digital Object Identifier 10.1109/JIOT.2018.2866164, online available: https://ieeexplore.ieee.org/document/8440029/}}
}

%
%

\markboth{IEEE Internet of Things Journal,~2018.}%
{Shell \MakeLowercase{\textit{et al.}}: Bare Demo of IEEEtran.cls for Journals}
%



\maketitle

\begin{abstract}
Recently, code reuse attacks (CRAs), such as return-oriented programming (ROP) and jump-oriented programming (JOP), have emerged as a new class of ingenious security threats. Attackers can utilize CRAs to hijack the control flow of programs to perform malicious actions without injecting any codes. Many defenses, classed into software-based and hardware-based, have been proposed. However, software-based methods are difficult to be deployed in practical systems due to high performance overhead. Hardware-based methods can reduce performance overhead but may require extending instruction set architectures (ISAs) and modifying the compiler or suffer the vulnerability of key leakage. To tackle these issues, this paper proposes a new hardware-assisted control flow checking method to resist CRAs with negligible performance overhead without extending ISAs, modifying the compiler or leaking the encryption/decryption key. The key technique involves two control flow checking mechanisms. The first one is the encrypted Hamming distances (EHDs) matching between the physical unclonable function (PUF) response and the return addresses, which prevents attackers from returning between gadgets so long as the PUF response is secret, thus resisting ROP attacks. The second one is the linear encryption/decryption operation (XOR) between the PUF response and the instructions at target addresses of call and jmp instructions to defeat JOP attacks. Advanced return-based full-function reuse attacks will be prevented with the dynamic key-updating method. Experimental evaluations on benchmarks demonstrate that the proposed method introduces negligible 0.95$\%$ runtime overhead and 0.78$\%$ binary size overhead on average.
\end{abstract}


\IEEEpeerreviewmaketitle

\section{Introduction}

\subsection{Motivation}
\IEEEPARstart{C}{ode} reuse attacks (CRAs) emerged as a powerful attack, can hijack the control flow of programs without injecting any malicious codes. CRAs can use the original code in the program to construct small fragments of existing codes. These code fragments are called gadgets. A gadget is composed of several assembly instructions, and each instruction can implement a different function (e.g., write a specified value to a fixed register). By chaining the gadgets ingeniously, an attacker can construct a sequence of gadgets to implement the same function as the malicious code. After constructing gadgets, attackers utilize the buffer overflow vulnerability to overwrite the return address on the stack with the start address of gadget chaining to hijack the control flow of the program, and ultimately obtain the system privilege.
CRAs mainly include return-oriented programming (ROP) attacks \cite{Checkoway2010,Snow2013} and jump-oriented programming (JOP) attacks \cite{Bletsch2011,Chen2011}. ROP was shown to be Turing-complete on a variety of platforms. It allows attackers to execute arbitrary codes by performing a chain of gadgets which come from existing binaries and all end with a ret instruction, while JOP makes use of jmp instructions instead of ret instructions in gadgets to change the control flow. Some tools\cite{Salwan2011} have been developed to find useful gadgets in given binaries to facilitate CRAs which have been demonstrated on a broad range of architectures, such as x86, Atmel AVR, SPARC, ARM, Z80 and PowerPC, and successfully cracked some well-known software such as Adobe Reader, Adobe Flashplayer and Quicktime Player.
To thwart the CRAs, many defenses have been proposed in industry and academia. However, they suffer from several issues. Below we will summarize these techniques and analyze the limitations of them.

\subsection{Limitations of Prior Art}
Buffer overflow is one of the most prevalent software attacks. Stack smashing \cite{Afek2007} is a typical buffer overflow attack that attackers inject the malicious code in the stack and overwrite the return address of the normal instruction to execute the malicious code. Several approaches have been proposed to defeat buffer overflow attacks. For example, data execution prevention (DEP)\cite{Team} prohibits a memory page from being writable and executable at the same time. Hence, attackers are unable to execute their injected malicious codes. DEP has been supported by both AMD and Intel processors and widely adopted by modern operating systems. However, CRAs are able to redirect the control flow of programs via reusing gadgets, thus eliminating the need of code injection and bypassing hardware memory protection mechanisms such as DEP.

Recently, a lot of defenses have been proposed to defend against CRAs. The most general solution is the control-flow integrity (CFI) checking, where the control flow graph (CFG) of the program is generated during compilation and enforced at runtime \cite{ Abadi2009}. CFI can be roughly classed into two categories, software-based and hardware-based. Current software-based solutions incur significant performance overhead \cite{ Abadi2009,Zhang2013,Kayaalp2012,Zhang2015a}, which limits adoption in practical systems. Hardware-based CFIs can reduce performance overhead and have attracted much attention in recent years. But hardware-based CFIs require extending the instruction set architectures (ISAs) of processors and modifying the compiler \cite{Budiu2006,Davi2014,Davi2015,Sullivan2016,Christoulakis2016,Qiu2017}, or suffer the vulnerability of leaking encryption/decryption key\cite{Qiu2016}.

\subsection{Our Contributions}
This paper proposes a Hardware-assisted Control-flow Integrity Checking technique, named HCIC, to thwart CRAs by encrypted Hamming distance (EHD)-matching and linear encryption/decryption without modifying compiler, extending ISAs or leaking encryption/decryption key. HCIC provides the following capabilities against ROP/JOP style control-flow anomaly attacks.

\textbf{EHD matching method:} the EHDs between return addresses of call-function and the physical unclonable function (PUF) \cite{Zhang2014} response are computed before the return addresses are stored in the memory structure. Then, the EHDs will be computed at runtime when the ret instructions are executed. Finally, the pre-computed EHDs are compared with the EHDs computed at runtime to verify whether the EHDs are matched. If an attacker modifies the return address in the stack, the EHDs will not be matched. In this case, our proposed mechanism can prevent the attacker from returning between gadgets to thwart ROP attacks. With the PUF response (key) updating, return-based full-function reuse attacks can be prevented.

\textbf{Linear encryption/decryption method:} the PUF-based linear encryption operations (XOR) on the instructions at target addresses of jmp and call are performed once the executable binary is loaded into memory. Then, the runtime linear decryption operation can be done while jmp or call instructions are executed. If an attacker modifies the destination address of a jmp or call instruction, the default decryption operation for the instruction at the address will be enforced, which will cause the control flow to be abnormal and eventually may result in a system error, thus defeating JOP attacks.

Compared with previous solutions, our solution incurs extremely low performance and binary size (memory) overheads. This approach has the following contributions and features:

\begin{enumerate}
  \item EHD matching method is proposed to protect programs from ROP attacks. This new method solves the security vulnerability suffered in the previous work\cite{Qiu2016} that the secret PUF key can be deduced through memory leakage or debugging.
  \item linear encryption/decryption mechanism is proposed to resist JOP attacks. In this mechanism, the PUF key used in the JOP-defense mechanism is different from the PUF key used in the ROP-defense mechanism, which brings a big advantage that the key leakage in JOP-defense mechanism cannot be used to compromise the ROP-defense mechanism. Therefore, the security is improved.
  \item Dynamic key-updating method is proposed to resist advanced CRAs and hence improves the security. Moreover, HCIC does not require the PUF to generate the reliable response and hence avoiding the intractable reliability issue suffered in previous PUF applications \cite{Koushanfar2012,Zhang2015b,Zhang2015c}.
  \item HCIC does not modify the compiler or add any new instructions to the existing processor's ISA.
  \item Simulation results demonstrate that our architecture's average runtime and binary size overheads are only 0.95$\%$ and 0.78$\%$, respectively, which are much lower than traditional CFI approaches (up to 21$\%$ \cite{ Abadi2009}).
  \item Exception analysis shows that our proposed defense does not produce anomalies when some exceptional cases occurred, and security analysis shows that the proposed method is sound and secure against CRAs with zero false positive and negative rates.
\end{enumerate}

The rest of this paper is organized as follows. Related work is elaborated in Section II. Code reuse attacks are described in Section III. Preliminaries are given in Section IV. The proposed hardware-assisted CFI and its working mechanism and exception analysis are elaborated in Section V. Security is analyzed in Section VI. The detailed experimental results and analysis are reported in Section VII. Finally, we conclude in Section VIII.

\section{Related Work}

CRAs use existing codes to launch an attack without injecting any codes. A kind of early code reuse attack, named return-to-libc attack \cite{Designer2007}, can call the libc functions elaborately chosen by the attackers to change the normal control flow. In this way, an attacker might eventually execute malicious computing. Recently, ROP \cite{Checkoway2010,Snow2013} and JOP\cite{Bletsch2011,Chen2011} as more powerful types of code reuse attacks are proposed. A lot of works have been proposed to defend against CRAs, such as address space layout randomization (ASLR) \cite{Bhatkar2005,Pappas2012,Wartell2012}, shadow stack \cite{Frantzen2001,Dang2015,Intel2017}, gadgets checking \cite{Chen2009} and CFI\cite{Davi2014,Davi2015,Sullivan2016,Christoulakis2016,Abadi2005}.

ASLR\cite{Bhatkar2005,Pappas2012,Wartell2012} is to randomize addresses of code and data regions to prevent the attacker from getting the entry address of gadgets when the program is loaded into memory. However, the data and code region are not fully randomized, and with the knowledge of some randomized codes, it is still possible for attackers to find enough gadgets in memory to perform CRAs \cite{Snow2013}. Shadow stack was proposed to protect the return addresses in the stack from being tampered by adding a second stack that is dedicated to control transfer operations. However, shadow stack is vulnerable to JOP attacks \cite{Frantzen2001,Dang2015}, and requires the specialized maintenance which brings additional overhead, and even requires modifying the ISA \cite{Intel2017}. Gadgets checking \cite{Chen2009} judges whether the program is attacked by monitoring the frequency of executing gadgets. This method can defend against JOP attacks. However, when a program consists of many small functions having little amount of instructions, it may incur a high false positive rate. Among current defenses, CFI is the most general solution whose key idea is to generate the CFG for a program during compilation and enforce the control flow to follow the CFG at runtime. CFI includes software-based CFI and hardware-based CFI as we will introduce in details next.

\subsection{Software-based CFI}
CFI checking makes the control flow change be consistent with the CFG of original program. Abadi et al. proposed to check the CFI \cite{ Abadi2009} by inserting the checking ID before each indirect branch instruction to prevent the unintended change of control flow. Any illegal change of control flow will be theoretically checked and rejected at runtime due to an ID-checking violation. In theory, each control flow transition can be inserted with a unique ID. However, these CFIs incur high performance overhead due to the ID creating, storing, querying and comparing. The performance overhead of CFI is up to 21$\%$ \cite{Abadi2009}. In order to reduce the performance overhead, several techniques \cite{Zhang2013} have been proposed to loosen the control transitions and use fewer IDs. Compact control flow integrity and randomization (CCFIR)\cite{Zhang2013} redirects indirect jmp branches to a new springboard. Indirect jmp instructions are checked and only allowed to transit control flow to the springboard entries by assigning aligned entry to direct and indirect branch targets. CCFIR uses three IDs to restrict control flow transitions. Two IDs are used to return to sensitive or insensitive functions, and one ID is used for call or indirect jmp instructions. This looser CFI allows the control flow to transfer to an address that does not exist in the CFG, which makes it be possible for attackers to launch an advanced ROP attack. Besides, CCFIR requires program allocating aligned springboard to ensure control flow integrity, which largely increases the space requirements.

As discussed above, software-based CFIs incur high performance and binary size overheads, and require the insertion of checking instructions or the creation of accompanying data structures such as a stack during the execution of program, which may overwrite registers or flags at runtime and cause programs behaving abnormally \cite{Abadi2009}.

\subsection{Hardware-Based CFI}
Hardware-based CFIs can reduce performance overhead and hence have attracted much attention recently. Several hardware-based CFIs have been proposed, such as Branch Regulation (BR)\cite{ Kayaalp2012} and hardware-assisted CFI\cite{Davi2014,Davi2015,Sullivan2016,Christoulakis2016}.

BR \cite{ Kayaalp2012} uses hardware support to monitor the control flow, in which the indirect jmp instruction is restricted to jump to its own function or the first instruction of other functions. BR also adds a shadow stack to record legal return addresses and check the return addresses before the functions return. To improve efficiency, BR uses cache to access the shadow stack. BR adds BR-annotation to indicate a function and restrict indirect branch. However, BR allows the jmp instruction to transfer the control flow inside the function, which can be used by an attacker to perform malicious attacks. Kanuparthi\emph{ et al}. \cite{Kanuparthi2010} proposed to use the dynamic trusted platform module (DTPM) to support runtime integrity checking of a program. The key idea is to verify the hash value of each trace in CFG to check the integrity of control flow. However, DTPM incurs high performance overhead and cannot detect the control flow anomalies between basic blocks (BBs). Davi \emph{et al}. \cite{Davi2014,Davi2015} proposed a hardware-assisted fine-grained CFI which adds new CPU instructions to ISAs. It assigns a different label to each function to ensure that an indirect call instruction must comply with new CPU instructions. Since the label for a call function is activated at call time and will be checked at return time, ret instructions can only return to the most recent callsite. However, the recognition of labels requires modifying the compiler and extending ISAs. Besides, runtime checking requires a label state memory to store function labels, which increases the space overhead. In 2016, Sullivan \emph{et al}. \cite{Sullivan2016} enhanced the hardware-assisted fine-grained CFI to ensure the forward-edge and backward-edge control flows follow the CFG. Such defense can prevent ROP, JOP, and full-function reuse. But it also requires extending ISAs and modifying the compiler and code size overhead is up to 13.5\%. In our previous hardware-assisted CFI architecture \cite{Qiu2016}, the return addresses of function-call are encrypted and will be decrypted with the simple XOR operation when the corresponding instructions are executed, which is vulnerable to the debugging attack because the linearly encrypted addresses can be got through memory leakage or debugging and hence secret PUF response can be deduced. Later, the authors \cite{Qiu2017} proposed to replace XOR with the AES integrated in Intel processors to improve security, but this technique also needs to expand the ISA (added new AES encryption and decryption instructions)\cite{Gueron2010}. Cryptographic CFI (CCFI) \cite{Mashtizadeh2015} also protects control flow elements with AES. However, CCFI is built on source codes and has limitations in performance overhead and defending against JOP attacks. Clercq \emph{et al.} \cite{Clercq2016} use cryptographic mechanisms to encrypt and decrypt the instructions with control flow dependent information to enforce CFI, which does not need to extend ISAs. However, it incurs unacceptable performance overhead (up to 110\%), and when the same function is called multiple times, the instruction decryption will be wrong which would incur high false positive rate.

To address these issues, this paper proposes a hardware-assisted CFI checking with the encrypted Hamming distance (EHD)-matching and linear encryption/decryption without modifying the compiler, extending ISAs and the vulnerability of leaking encryption/decryption key. Our proposed method incurs low performance overhead and does not produce anomalies when some exceptional cases occurred. Security analysis shows that the proposed method is sound and secure against ROP and JOP and even some advanced CRAs such as return-based full-function reuse attacks.

\section{Code Reuse Attacks}
When executing a call function, CPU pushes the return address into the stack and then performs the instruction at the first address of the destination function. When the ret instruction is executed, the CPU pops the return address from the stack and executes the instruction at the return address. ROP attacks first construct a sequence of gadgets from the existing code, then link the entry addresses of these gadgets together to form a chain. Finally, an attacker exploits the buffer overflow vulnerability to overwrite the return address on the stack with the entry address of the gadgets chain. Once CPU executes the ret instruction, the program would execute the gadgets chain and completes the ROP attack. The principle of JOP is similar to ROP. The difference is that JOP uses indirect jmp instructions to hijack the program's control flow and execute malicious actions. In what follows, we will discuss ROP and JOP in detail.

\subsection{ROP attacks}

Stack is used to store temporary variables, the return address and parameters of the function call and so on, which will be pushed into the stack successively when CPU executes the call instruction. When the ret instruction is executed, the return address will be popped into the program counter and CPU will execute the instruction at the return address. If an attacker overwrites the return address, the instruction at the overwritten address would be executed, and the control flow will be hijacked by the attacker. ROP utilizes the characteristics of the stack for call function to hijack the control flow of programs by overwriting the return address in the stack. Gadgets are the object program's small code fragments, so the ROP attack is covert and difficult to be found. Such attack can bypass existing defense mechanisms such as DEP\cite{Team}.

\begin{figure}[!t]
  \centering
  \includegraphics[width=\linewidth]{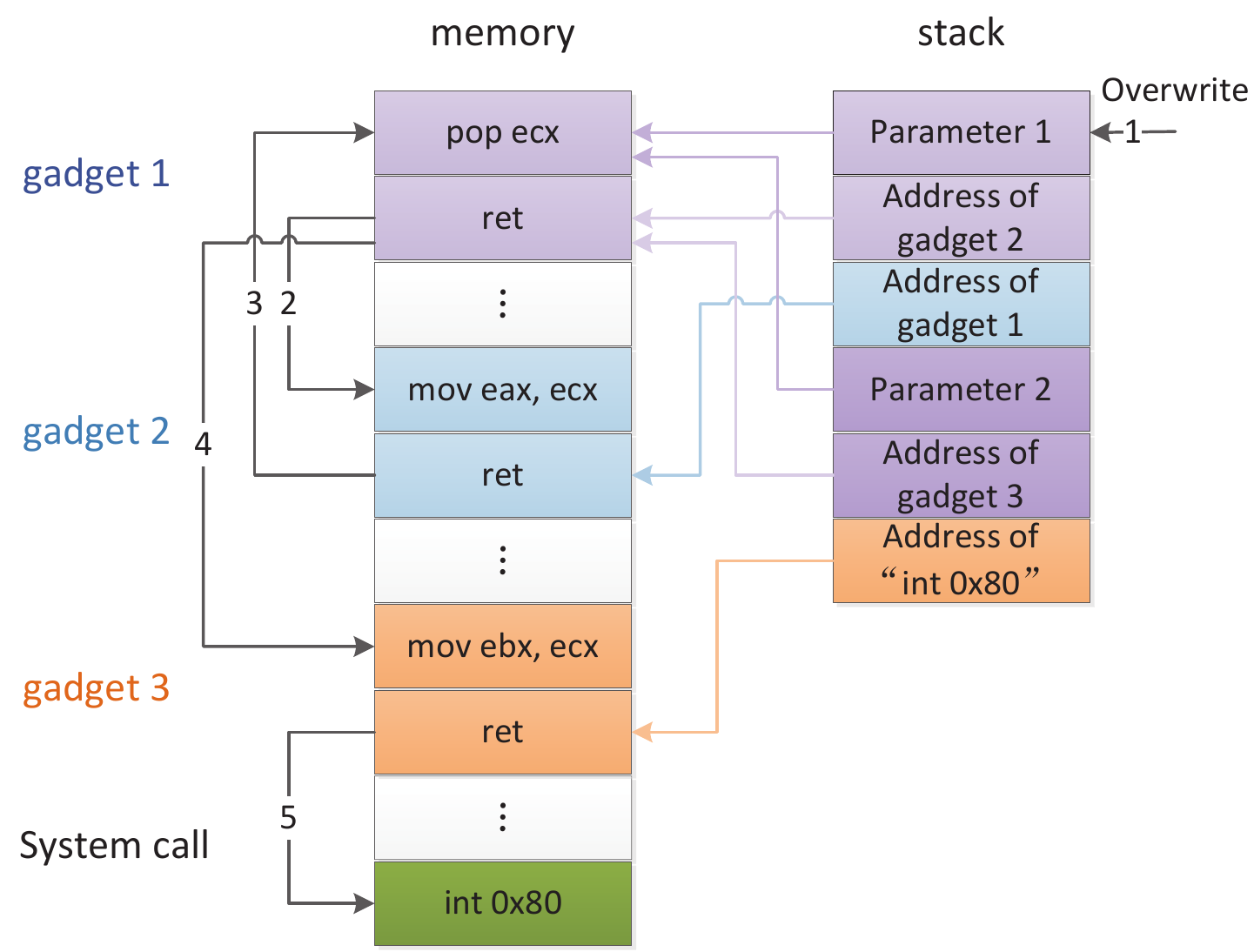}
  \caption{An example of ROP attack.}
  \label{fig1}
\end{figure}

We illustrate an ROP attack in Fig. \ref{fig1}, where the goal of the attacker is to make a system call ``int 0x80'' with the parameter ``data'' in both eax and ebx registers. To achieve this goal, the attacker needs to complete the following steps.

\textbf{Step 1}: The attacker exploits an overflow vulnerability to tamper the return address with the addresses of the system call ``int 0x80'', gadget 1, gadget 2, gadget 3 and the data that the system call needs in the stack.

\textbf{Step 2}: The program executes the gadget 1. The ``pop ecx'' instruction stores the parameter 1 which is designated by the attacker in ecx register. Then, the program gets the address of gadget 2 from the stack and jumps to the gadget 2 after executing the ret instruction.

\textbf{Step 3}: The program executes the gadget 2. The ``mov eax, ecx'' instruction moves the parameter 1 in the ecx to the eax. Then, the program gets the address of gadget 1 from the stack and jumps to the gadget 1 after executing the ret instruction.

\textbf{Step 4}: The program executes the gadget 1 again. The ``pop ecx'' instruction stores the parameter 2 which is designated by the attacker in the ecx register. Then, the program gets the address of gadget 3 from the stack and jumps to the gadget 3 after executing the ret instruction.

\textbf{Step 5}: The program executes the gadget 3. The ``mov ebx, ecx'' instruction moves the parameter 2 in the ecx to the ebx. Then, the program gets the address of ``int 0x80'' from the stack and jumps to ``int 0x80'' after executing the ret instruction.

When constructing the entry address chain of gadgets, we add the same number of data as the number of pop instructions after adding the entry address of gadgets. This ensures that when the ret instruction is executed, the program pops the return address which is the entry address of the next gadget and continues to execute the rest of gadgets to complete the attack.

\subsection{JOP attacks}
The JOP attack is similar to the ROP attack. It also utilizes the existing code in the program to hijack the program control flow. The difference is that JOP attacks use indirect jmp instructions to change the control flow of the program while the ROP attacks use ret instructions. During program execution, an attacker can change the values in registers with specified parameters. When program executes the indirect jmp instruction or indirect call instruction, the target address taken from the register is the address that the attacker constructed. The attacker forces the program to execute these gadgets to complete the JOP attack. Because JOP uses indirect jmp/call instructions instead of ret instructions, which makes current ROP defenses unable to prevent JOP attacks. Therefore, ROP and JOP should be prevented simultaneously for any effective defenses.

\section{PRELIMINARIES}
The general terms and concepts used throughout the paper will be introduced as follows.

\subsection{Silicon Physical Unclonable Functions}
Silicon Physical Unclonable Function (PUF) has emerged as a promising hardware security primitive that is used for authentication and key generation without the requirement of expensive hardware such as secure EEPROMs and battery-backed SRAM, and hence gained a lot of attention over the past few years \cite{Zhang2014}\cite{Herder2014}.

There are several reasons that we use PUF instead of traditional secret key storing in digital memory to assist CFI verification. First, PUFs derive a secret from the physical characteristics of the IC. This approach is advantageous over standard secure digital storage such as more easy to fabricate, consuming less power and area, and  naturally anti-tamper\cite{Herder2014}. Second, the PUF key is chip-unique, unclonable and can be updated each time when the program is loaded. Even if a PUF key is cracked in a system, it cannot be used to break another system, and hence improves the security largely. Third, it is well known that current PUF applications such as key generation \cite{Suh2007}, two-factor authentication \cite{zhang2018}, anti-overbuilding \cite{Koushanfar2012}, FPGA IP protection \cite{Zhang2015b,Zhang2016b} and resisting FPGA replay attacks \cite{Zhang2015c} cannot tolerate bit errors (e.g., key generation has an extremely high requirement on reliability). They all require extra reliability-enhancing techniques \cite{Pang2017} and error-correction schemes \cite{Majzoobi2012} to increase the quality of responses, which incurs significant hardware overhead and hence limits adoption. Our proposed HCIC exhibits a big advantage that it does not require the PUF to generate the reliable response and hence avoiding such intractable issue suffered in previous applications. Last, many intrinsic PUFs \cite{Tehranipoor2015,Maiti2012} have been proposed. They require no additional circuitry (zero hardware overhead), and is cost effective. Random number generator also can be used, but it incus higher hardware overhead and lower security than PUF.

\subsection{Control Flow Integrity}
Before running the program, the control flow integrity (CFI) mechanism calculates the normal execution paths of the program. CFI employs the debug information to generate the complete CFG before the program is compiled, and forces the program to execute in accordance with the normal control flow. To ensure CFI, the CFG needs to be extracted from the program first. A basic block (BB) is an instruction sequence that only has a single entry point and a single exit point. A function consists of multiple BBs. A CFG represents the correct run direction of program between BBs.

\begin{figure}
\centerline{\includegraphics[width=\linewidth]{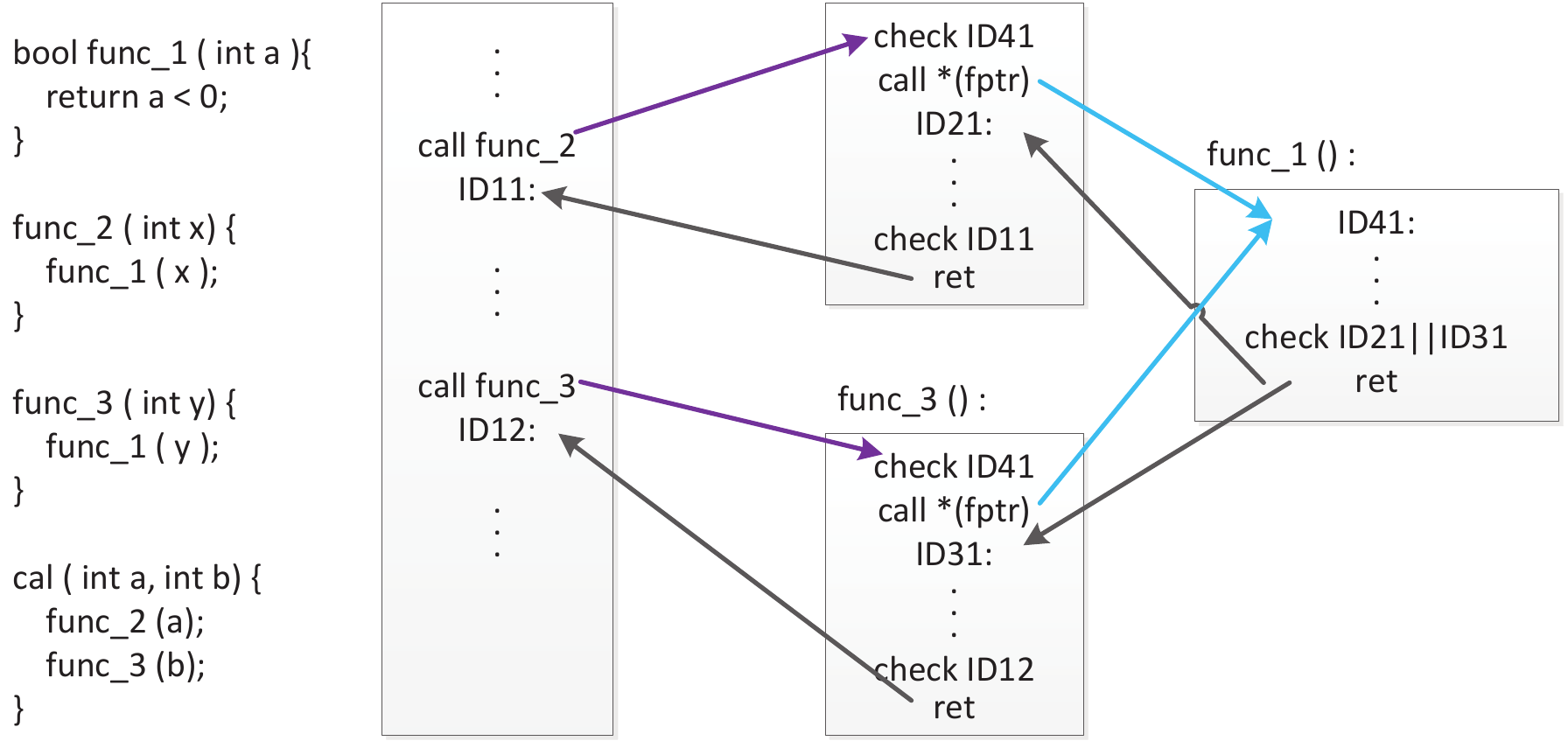}}
\caption{An example of program fragments and a fine-grained CFI.}
\label{fig2}
\end{figure}

As shown in Fig. \ref{fig2}, the fine-grained CFI \cite{Goktas2014} assigns a unique ID at the target address of control flow instructions and inserts the ID checking instructions before the control flow instructions in order to ensure that the target addresses of the control flow instructions are valid. There are two types of control flow jump instructions, direct jump and indirect jump. The target address of the direct jump is fixed and cannot be tampered, so the ID does not need to be inserted and checked when the program is running. While for indirect jump instructions, their target addresses are calculated and loaded into the memory when the program is running and hence can be tampered. The CFI is for indirect jump instructions. Before the program executes the indirect jump instruction, it is first to check that whether the ID at the target address is equal to the known and valid ID of the jump instruction. With this way, the legitimacy of indirect jump is verified. However, the fine-grained CFI inserts a unique ID at the target address of indirect jump. If the instruction can jump to multiple target addresses, all these IDs will be compared to determine whether the jump is legal. This incurs high performance overhead.

\begin{figure}
\centerline{\includegraphics[width=\linewidth]{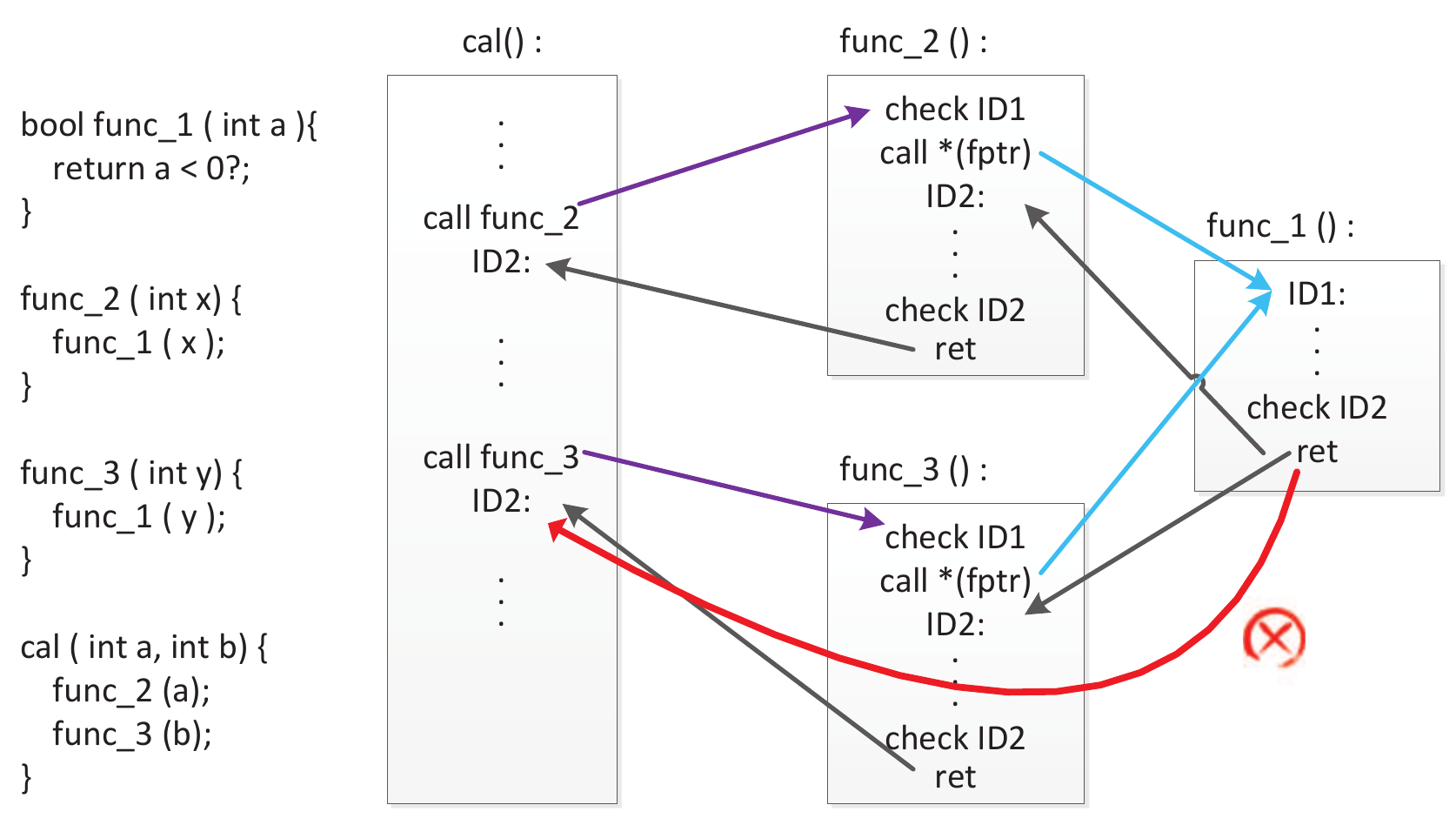}}
\caption{An example of program fragments and a coarse-grained CFI.}
\label{fig3}
\end{figure}

Fig. \ref{fig3} shows an example of a coarse-grained CFI with fewer IDs to reduce the performance overhead, much like CCFIR \cite{Zhang2013} and Bin-CFI \cite{Zhang2015a}. In this example, there are two kinds of control flow transitions. The first one is the function pointer fptr pointing to function func\_1() (in function func\_2() and  func\_3()), and the other is the function return of func\_2(), func\_3(), cal(), or func\_1(). In order to prevent ROP and JOP attacks from tampering the destination address of fptr, CFI adds the ID1 checking (check ID1) for verifying legal jumps from the source (function func\_2() and func\_3()) to the destinations of fptr (functions func\_1()). Any illegal jumps to other destinations will not pass the check ID1 because there are no ID1 inserted at those destination addresses. Meanwhile, CFI also adds the ID2 checking (check ID2) for verifying legal returns from the call functions to the callsite to avoid malicious modifications on the return addresses in the stack. Any changes to other return addresses will not pass the check ID2 because there are no ID2 inserted at illegal addresses. This CFI introduces fewer IDs (for instance, the Bin-CFI mechanism introduces two IDs). Therefore, it introduces lower verification overhead than the ideal CFI. However, as the gadgets with the same ID increases, the attack would be success with higher probability. For example, the jump from func\_1 to func\_3 (red) shown in Fig. \ref{fig3} is illegal but cannot be detected with this CFI. Recently, several coarse-grained CFI methods have been proposed in academia and industry. For example, Intel proposed an indirect branch tracking method that adds a new instruction ENDBRANCH to mark valid indirect call/jmp targets in the program to defend against JOP \cite{Intel2017}; BBB-CFI \cite{He2017} proposed to limit the indirect call/jmp to target at the starting address of a BB by code-inspired BB boundary and data-inspired BB boundary to defend against JOP. These coarse-grained methods allow the indirect call/jmp to target at the starting address of any BBs and hence cannot check the unintended branches between BBs.

As described above, fine-grained CFI and coarse-grained CFI have their own strengths and weaknesses. The fine-grained CFI has higher security, while the coarse-grained CFI incurs lower performance overhead. Therefore, the security and practicality should be balanced. Our proposed method in this paper shows a good balance between security and practicality compared with previous CFI methods because HCIC incurs low performance overhead and is able to resist mainstream CRAs.

\section{The Proposed Hardware-assisted CFI}
\begin{figure}
\centerline{\includegraphics[width=\linewidth]{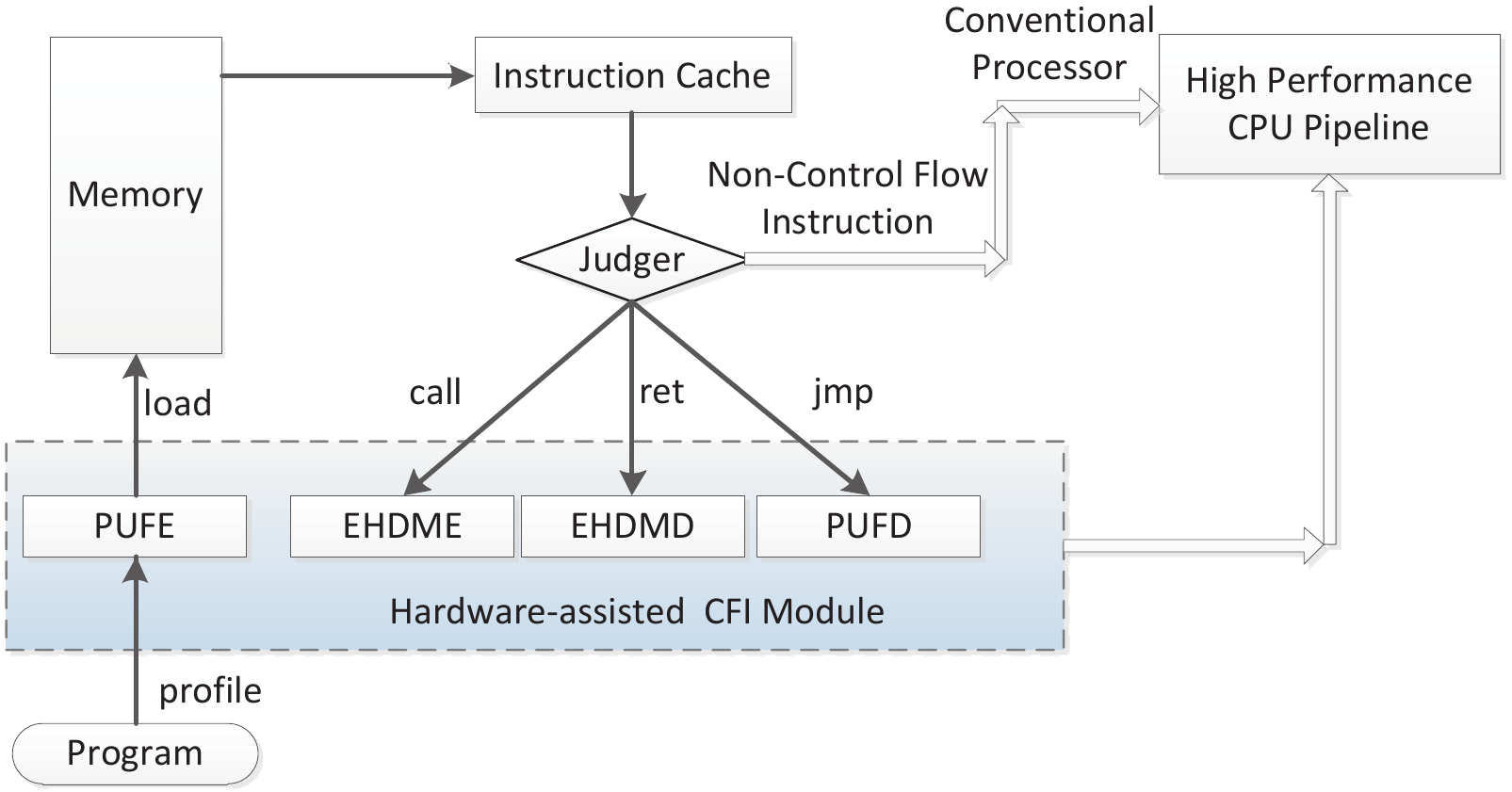}}
\caption{Basic framework of HCIC}
\label{fig4}
\end{figure}

In this study, we assume that the program code is trustworthy and the attacker cannot modify the source code of the program that has been loaded into memory. In addition, we assume that the attacked system has deployed DEP which forces the attacker to reuse existing code. We further assume that the attacker can access the stack to perform overflow attacks and CRAs. The details of the proposed hardware-assisted CFI checking are depicted in figures \ref{fig4}, \ref{fig5} and \ref{fig6} and described as follows.

\subsection{Encrypted Hamming Distance}
Hamming distance (HD) is a number used to denote the difference between two binary strings. Let $x$ and $y$ be two binary sequences of the same length. The HD $d(x, y)$ between $x$ and $y$ is the number of positions at which the corresponding symbols are different:

\begin{equation}\label{(1)}
d(x, y)=\sum_{i=0}^{n-1}x[i]\bigoplus y[i]
\end{equation}
where \emph{i} = 0, 1, ..., \emph{n}-1; \emph{x}, \emph{y} denotes \emph{n}-bit binary sequence;  $\oplus$ denotes exclusive OR.

One of key contributions in this paper is that the encrypted Hamming distance (EHD) matching method is proposed to resist ROP. The EHD is generated by appending random $l$ bits at the end of $d(x, y)$ and then rotating right \emph{m} bits. In this paper, $m$ can be the last $k$-bit in the PUF key (ranges from 0 to $2^k-1$); $l$ can be the first $l$-bit in the PUF key ($l = 2^k-6$). Since the address is 32 bits, the HD ranges from 0 to 32, and requires 6 bits to encode.

For example, assume that the secret PUF key $key\_2$ is 0x12345678 (0001 0010 0011 0100 0101 0110 0111 1000), HD = 20 (010100), $k$ = 5, and $l$ = 26. First, we append the first 26 bits of $key\_2$ (0001 0010 0011 0100 0101 0110 01) to the end of the HD, and then rotate the new 32-bit HD (0101 0000 0100 1000 1101 0001 0101 1001) right $m$ bits. The value of $m$ is determined by the last 5 bits of $key\_2$ (11000), so $m$ = 24. Finally, the 32-bit HD is rotated right 24 bits to generate the EHD 0x48D15950 (0100 1000 1101 0001 0101 1001 0101 0000).

\subsection{Framework of HCIC}

\begin{figure}
\centerline{\includegraphics[width=\linewidth]{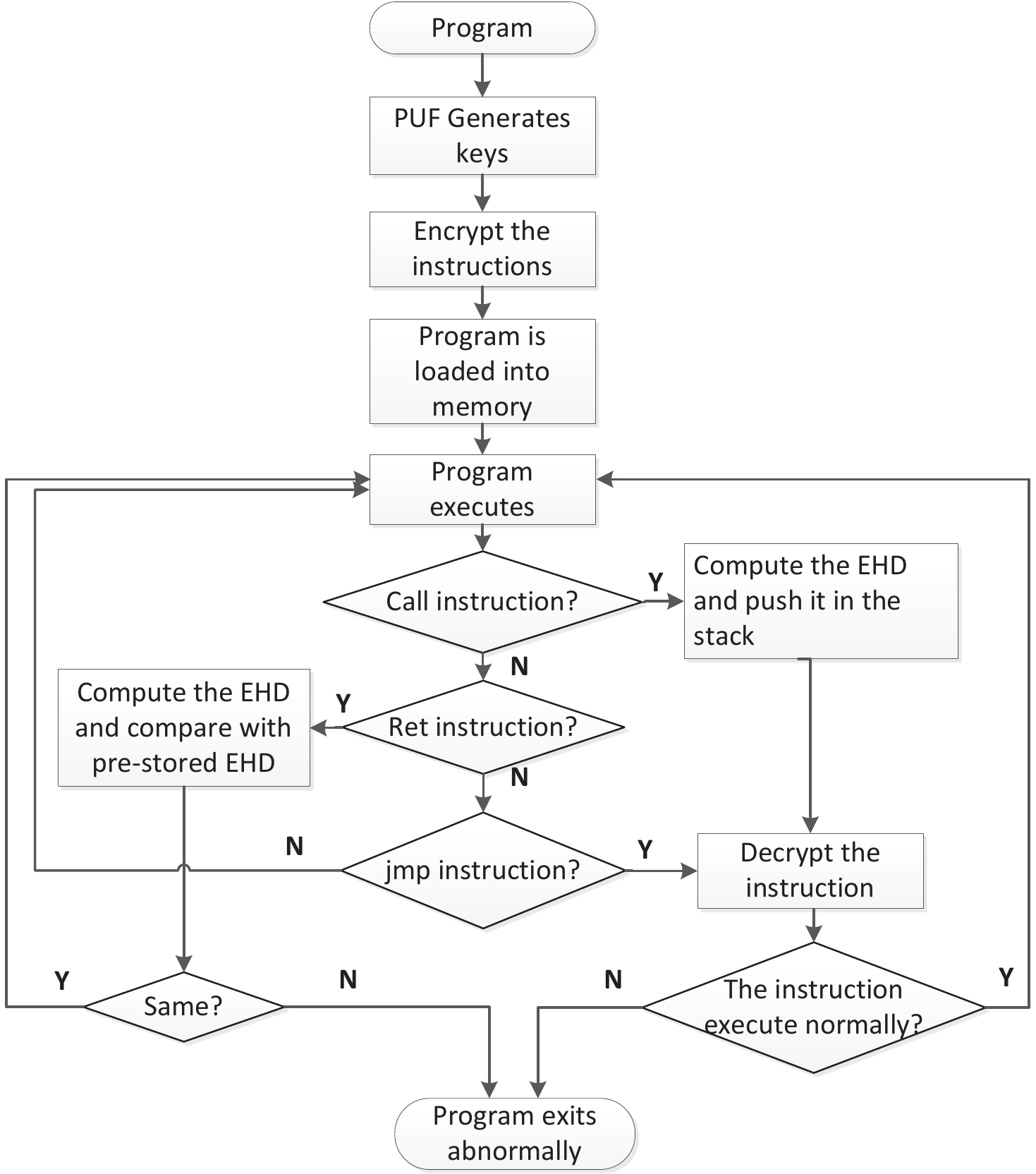}}
\caption{The flow of HCIC}
\label{fig5}
\end{figure}

The framework of hardware-assisted CFI checking is shown in Fig. \ref{fig4}. Judger is to determine whether the instruction is the control flow instruction. If the fetched instruction is the control flow instruction, the next instruction would be processed by the hardware-assisted CFI Module, while non-control flow instruction is fed to the conventional processor. There are four key operations, EHD matching-based encoding (EHDME), EHD matching-based decoding (EHDMD), PUF-based encoding (PUFE) and PUF-based decoding (PUFD), involving in our proposed HCIC. EHDME and EHDMD are used to encode and verify the EHDs to resist ROP attacks. PUFE and PUFD are used to resist JOP attacks. The whole flow of HCIC is shown in Fig. \ref{fig5}.

A ret instruction will bring the program execution to an address which is pushed into the stack by a call instruction. But an attacker is able to modify the address by overflow attacks. We should guarantee the ret instruction targeting the next instruction of the corresponding call instruction. The jmp instructions and call instructions also should be limited to point to the encrypted instructions which belong to the first instructions of BBs. We elaborate it as follows.

\textbf{CALL and RET}: To prevent attackers from changing the return address of a ret instruction via the memory overflow and then hijacking the control flow of a program, EHDME would first compute the encrypted Hamming distance EHD1 between the $key\_2$ generated by PUF and the return address when the call instruction is executed. Then, the return address and EHD1 are pushed into the stack. When the ret instruction is executed, EHDMD would compute the encrypted Hamming distance EHD2 between $key\_2$ and the current popped return address again. Only the EHD2 is equal to the pre-stored encrypted Hamming distance EHD1, can the program jump normally. Once the return address was overflowed at runtime or there is no paired call for ret instructions, the EHD2 would be unequal to EHD1 because the PUF key is unknown to the attacker. Hence, the execution cannot point to the code that attackers intend to perform.

\textbf{CALL and JMP}: We use the XOR operation to linearly encrypt/decrypt the first instruction at destination address of the jmp and call in order to prevent attackers from hijacking the control flow of a program. Before the program is loaded into memory, the $key\_1$ generated by PUF is used to encrypt all first instructions at target addresses (PUFE). When the program is running, as long as the jmp or call instruction is executed, the first instruction will be decrypted dynamically with $key\_1$ (PUFD). Thus, if attackers hijack the control flow via illegally changing the target address of the jmp and call, the program may jump to an address with an unencrypted instruction which would be decrypted forcedly. If an incorrect decrypted instruction is executed, a system error may occur. Thus, JOP attacks would be defeated if an attacker wants to use indirect branch instructions to perform code reuse attacks.

\begin{figure}
\centerline{\includegraphics[width=\linewidth]{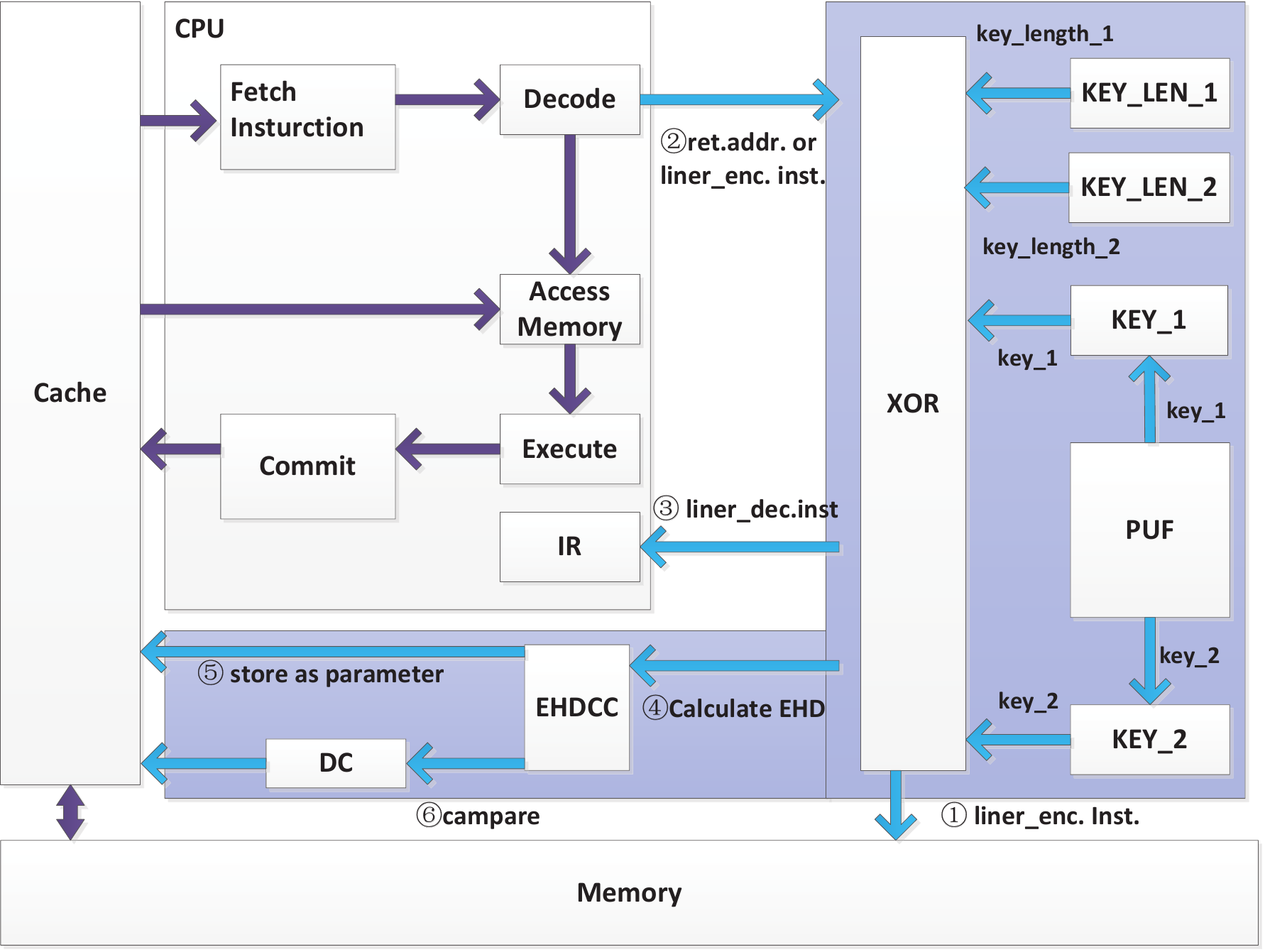}}
\caption{Micro-architecture and data flow of HCIC. \textcircled 1 $key\_1$ is used to linearly encrypt the first instruction at target address of jmp and call via XOR operation. \textcircled 2 Perform XOR operation on the return address of the call instruction with $key\_2$ or the first instruction at target address of the jmp instruction and the call instruction with $key\_1$. \textcircled 3 $key\_1$ is used to linearly decrypt the first instruction at target address of the jmp and call instruction via XOR operation. \textcircled 4 Calculate the encrypted Hamming distance EHD1 between the return address and $key\_2$. \textcircled 5 When a call instruction is executed, EHD1 is pushed into the stack as a parameter. \textcircled 6 When a ret instruction is executed, the encrypted Hamming distance EHD2 between $key\_2$ and the current popped return address is calculated and transmitted into the decision circuit (DC) which judges whether EHD2 is equal to the pre-stored EHD1.}
\label{fig6}
\end{figure}

 Fig. \ref{fig6} depicts the hardware structure of our proposed mechanism, as shown in the purple area, and how it interacts with CPU. The hardware structure is appended to CPU architecture with a XOR operation unit, 4 registers ($KEY\_1, KEY\_2, KEY\_LEN\_1, KEY\_LEN\_2$), a PUF module, an encrypted Hamming distance calculation circuit (EHDCC) and a decision circuit (DC). CPU communicates with the XOR unit; the PUF communicates with the registers; the registers communicate with the XOR unit. The XOR unit communicates with EHDCC; the EHDCC communicates with the DC. $key\_1$ and $key\_2$ used in XOR operation are generated by the PUF module and stored in $KEY\_1$ and $KEY\_2$ registers. The reason we use two keys is to solve the security vulnerability that the work \cite{Qiu2016} suffered. It is well-known that if the program has been loaded into memory, the code can not be tampered anymore. Hence, JOP attacks are unable to break the defense. However, through JOP attacks, the attacker can deduce the secret key used for linear encryption by memory leakage or debugging. If the defense mechanism exploits a single secret key in all XOR operations, attackers can calculate correct EHDs of every address for ROP attacks. To eliminate the vulnerability, we exploit two different secret keys for the encryption/decryption operation. The key length of $key\_1$ and $key\_2$ is determined by the $KEY\_LEN\_1$ and $KEY\_LEN\_2$ registers. Here the length of $key\_1$ is $x$ which is less than the length of the shortest first instruction, and the length of $key\_2$ is 32 bits. The values of $KEY\_LEN\_1$ and $KEY\_LEN\_2$ registers are $x$ and 32, respectively. After receiving the XOR result, the encrypted Hamming distance of current return address is calculated in EHDCC and then is sent to DC. The DC will judge whether the calculated encrypted Hamming distance is equal to the previously stored one.

In our proposed HCIC, the first step is to perform XOR operation on all first instructions at target addresses of call and jmp instructions for linear encryption before the program is loaded into memory. Next, when the jmp instruction is executed, the first instruction at target address will be decrypted dynamically. When the call instruction is executed, XOR and EHDCC will be used to calculate the encrypted Hamming distance between the return address and $key\_2$ and store it in stack, and then the first instruction at target address is decrypted and executed. When the ret instruction is executed, the XOR and EHDCC will be used to calculate the encrypted Hamming distance between return address and $key\_2$, and then the DC judges whether the program control flow is normal by comparing the calculated EHD and the previously stored one.

The concrete process of HCIC is as follows:

\begin{enumerate}
  \item When the program is started up, CPU calls the silicon-PUF module to generate $key\_1$ and $key\_2$ and stores them in $KEY\_1$ and $KEY\_2$ registers, respectively. According to the rule proposed above, the lengths of $key\_1$ and $key\_2$ are generated and stored in $ KEY\_LEN\_1$ and $KEY\_LEN\_2$, respectively.
  \item Before loading the program into memory, CPU searches target addresses of all call and jmp instructions and transmits every first instruction to XOR.
  \item XOR unit performs XOR operation on every first instruction for linear encryption unit according to $key\_length\_1$ and $key\_1$ and returns the result to CPU.
  \item After all the first instructions are encrypted, the program is loaded into memory and starts up.
  \item In the executing process, memory transmits the program instruction to Cache in real time. CPU will fetch the instruction from Cache and then decrypt it. When the call instruction is fetched, jump to the step 6) to perform; when the ret instruction is fetched, jump to the step 9) to perform; when the jmp instruction is fetched, jump to the step 13) to perform; if the program is done, exit normally.
  \item Transmit the return address of the call instruction to XOR unit.
  \item XOR unit performs XOR operation on return address with $key\_length\_2$ and $key\_2$, and sends the result to EHDCC.
  \item EHDCC calculates the encrypted hamming distance between the current return address and $key\_2$ according to the received result. Then the encrypted Hamming distance is pushed into the stack as a parameter of the call instruction, following the return address. Jump to the step 13) to perform.
  \item Transmit the return address of the ret instruction to XOR unit.
  \item XOR unit performs XOR operation on the return address with $key\_length\_2$ and $key\_2$, and sends the result to EHDCC.
  \item EHDCC calculates the EHD between the current return address and $key\_2$ according to the received result and sends the encrypted Hamming distance to DC.
  \item DC unit receives the EHD from EHDCC and compares it with pre-stored EHD. If the comparing result is the same, program continues to perform, and jump to the step 5); if not, jump to the step 16).
  \item Program jumps to target address of the call or jmp instruction. When the instruction at target address is loaded in CPU, the content of IR register is transmitted to XOR unit.
  \item XOR unit performs XOR operation toward the instruction with $key\_length\_1$ and $key\_1$ for linear decryption and returns the result to IR register of CPU.
  \item CPU executes the instruction which has been decrypted within IR register. If the decrypted instruction is executed normally, jumps to 5); if not, the program exits abnormally.
  \item The attack is detected; system produces warning and waits to be processed.
\end{enumerate}

\begin{figure}
\centerline{\includegraphics[width=\linewidth]{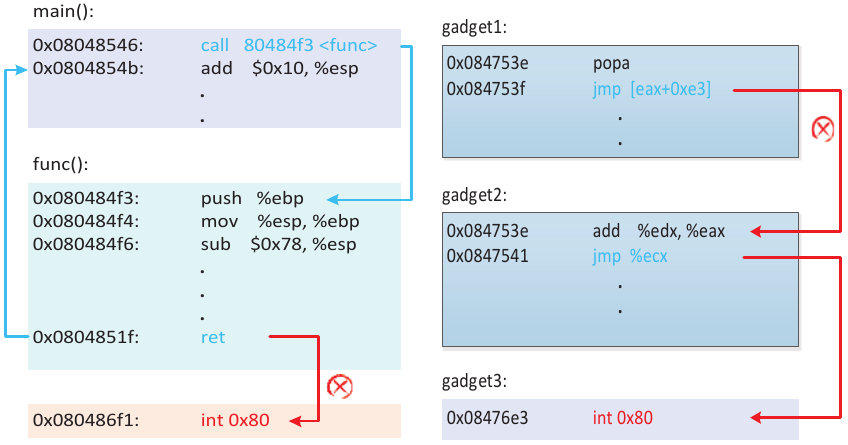}}
\caption{Examples to resist CRAs. a) An example to resist ROP attacks. b) An example to resist JOP attacks.}
\label{fig7}
\end{figure}

In what follows, we use a 32-bit CISC instruction architecture to illustrate the key idea of our approach. In this case, a \emph{n}-bit PUF is used to encode \emph{n}-bit instructions at the target address of the call and jmp instructions and compute the EHD. We give two examples to demonstrate how code-reuse attacks would fail in attacking programs protected by the HCIC. As an example, as shown in Fig. \ref{fig7}a), a call instruction diverts execution to the entry point of a function $func()$. Assume the return address $A1$ is 0x0804854b and $key\_2$ is 0xa2156cf7. The HD between $A1$ and $key\_2$ is 16 (010000). `01000010' is generated by padding the first 2-bit `10' of $key\_2$ to `010000'. Then `01000010' is rotated right 7 bits (because the last 3-bit `111' of $key\_2$ is equal to 7) to get the encrypted Hamming distance EHD1 132 (10000100) which is pushed into the stack. If an attacker has overflowed the return address in order to perform the instruction ``int 0x80'' at the address $A2$ (0x080486f1), the overflowed address $A2$ will be fetched when ret is executed. However, the encrypted Hamming distance EHD2 between $A2$ and $key\_2$ is computed with XOR and EHDCC. DC determines whether EHD1 is equal to EHD2. Because the attacker does not know the PUF key, EHD1 (132) is not equal to EHD2 (108). In this case, the program would throw an exception and generate a warning. Hence, the program will not perform the intended instruction ``int 0x80''.

As another example, as shown in Fig. \ref{fig7}b), we assume an attacker overflows the program successfully to perform JOP attacks. When the jmp in gadget1 is executed, the program would be deviated to the gadget2. However, the XOR would be used to decrypt the target instruction ``add \%edx, \%eax''. The decryption would be wrong for the target instruction because this unintended instruction did not be encoded by XOR operation. Hence, our proposed HCIC can efficiently thwart the code reuse attacks.

\subsection{Exception Analysis}
In complex binaries, there are some exceptions to the normal behavior of call, ret, and jmp instructions. For example, most of the previously proposed CFI methods that strictly follow call-ret pairing would result in false positives even for the normal control flow of the program when call-ret pair is replaced by call-jmp pair. In this section, we will discuss exceptional cases in detail, and prove that our proposed defense does not produce these anomalies.

\emph{Case 1: the jmp instruction replaces the ret instruction to target the return address.}

There are two cases that the jmp instruction will replace the ret instruction to target the return address: 1) the $longjmp()$ function in the C language. After $longjmp()$ function is executed, the indirect jmp instruction targets the return address instead of the ret instruction; 2) the compiler sometimes replaces ret instructions with the $pop$ and $jmp$ instructions. In these cases, a CFI policy which strictly follows call-ret pairing would result in false positives.

HCIC does not result in false positives in these cases. When the call instruction is executed, the return address of the function is pushed into the stack with the EHD. When the function is done, the jmp instruction instead of the ret instruction will be executed to target the return address. In addition, the first instruction at the target address of the jmp instruction has been encrypted before the program is loaded into memory. Therefore, when the jmp instruction instead of the ret instruction is executed, the program will decrypt and execute the first instruction at the target address successfully. The stack space of the function will be reclaimed after the function is done.

\emph{Case 2: call-ret instructions appear in different order}

For the shadow stack technique, all threads share the same shadow stack. In a multithreaded program, thread \emph{A} executes the call instruction, and the return address \emph{R1} is pushed into the shadow stack. Then thread \emph{B} executes the call instruction, and the return address \emph{R2} is pushed into the shadow stack successively. In this case, if thread \emph{A} executes the ret instruction before \emph{B}, \emph{R1} will be popped to compare with the \emph{R2} at the top of the shadow stack, which would throw an exception due to the mismatch. In order to handle this exception, the return address is allowed to match all addresses in the shadow stack, which increases the performance overhead greatly.

In our proposed method, the exception does not be produced in the multi-threaded program because all call functions have their own stack and the EHDs between the PUF key and their return addresses are stored in the stack without affecting each other.

\emph{Case 3: ret instruction returns to a non-return address}

In the C++ exception handling mechanism, an exception is thrown when the program executes the \emph{throw}() function. After the \emph{throw}() is done, the program will jump directly with the ret instruction to the end of the try block instead of the next instruction after the \emph{throw}(). In this case, some defenses \cite{Das2016a} will produce an exception since the return address is not the return address stored in the shadow stack when the call instruction is executed. In order to handle the exceptional case, \cite{Das2016a} extended rules by adding the exception handler landing pad addresses in the $.eh\_frame$ and $.gcc\_except\_table$ to the shadow stack so that the return address of the ret instruction in \emph{throw}() can be matched in the shadow stack. However, in our experiments, we found that the \emph{throw}() function eventually jumps to the end of the try block by the \emph{jmp} instruction instead of the \emph{ret} instruction.

HCIC does not produce the exception. Similar to the case 1, when \emph{throw}() function is executed, the return address of the function is pushed into the stack with the EHD. When the \emph{throw}() function is done, the jmp instruction is executed to target the end of the try block. In HCIC, the first instruction at the target address of the jmp instruction has been encrypted before the program is loaded into memory. When the jmp instruction is executed, the program will decrypt and execute the first instruction at the target address successfully. Therefore, the \emph{throw}() function can be executed normally with the jmp instruction. The stack space of the function will be reclaimed after the function is done.

\section{Security Analysis}
For code reuse attacks, attackers first make use of the stack overflow to overwrite the return address to execute the first gadget, and then the ret/jmp/call instructions of gadgets are used to change the control flow between gadgets, which finally enables the attackers to execute malicious actions. Stack smashing and ROP attacks are based on overwriting of the return address. Our technique secures the target address of the ret instruction by the EHDME and EHDMD operations. Therefore, these attacks can be prevented. JOP attacks use the target address of the indirect jmp and call instruction which is limited to point to the first instructions of BBs. Since a part of first instructions of BBs have been encoded by the PUFE operation before the executable binary is loaded into memory, if any call and jmp instructions illegally change the control flow to the instruction which is not the first instruction of the BBs, the instruction would be wrongly decoded by the PUFD operation. With the PUF key updating, even the illegal control flow between BBs can be prevented. Thus, all the common attacks to the control flow can be prevented by our method. In what follows, the four important security threats or metrics are analyzed.

\subsection{The key leakage issue}
In \cite{Qiu2016}, the simple XOR operation is used to encrypt the return address in the stack and the instructions at the target addresses of the indirect jmp instructions. This scheme suffers the following key leakage issues:
\begin{enumerate}
  \item Attackers can get the original and encrypted return address by debugging and then deduce the secret key through the XOR operation.
  \item Attackers can get the original and encrypted instruction by debugging and then deduce the secret key through the XOR operation.
\end{enumerate}

HCIC is to calculate and compare the EHD between the return address and the secret key to resist the ROP attack, and decrypt the instruction at the target address of jmp and call instructions to resist the JOP attack. Although the return address and the EHD between the return address and the secret PUF key are pushed into the stack, the attacker cannot deduce the key with the EHD and the address. Therefore, HCIC resolves the above key leakage issue and hence exhibits the higher security.

In addition, HCIC uses two registers $KEY\_1$ and $KEY\_2$ to store different PUF keys against JOP and ROP, respectively, which further improves the security. If the EHD calculation and the instruction encryption use the same key, attackers can get the key $key\_1$ used for JOP defense by debugging, and then use $key\_1$ to crack the ROP defense. With different PUF keys for EHD calculation and instruction encryption, even if the attacker can obtain the $key\_1$ in the JOP defense by debugging, $key\_1$ cannot be used for ROP attacks. Moreover, the $key\_1$ cannot be used to launch JOP attacks anymore because HCIC will encrypt the instructions before the program is loaded into memory and decrypt the instructions dynamically when the program is running (It is well-known that if the program has been loaded into memory, the code can not be tampered anymore). Therefore, HCIC does not suffer the key leakage issues.

\subsection{EHD guessing attacks}
In our proposed ROP defense, if attackers guess the correct encrypted Hamming distance (EHD) between the return address and the secret PUF key $key\_2$, the defense would be compromised.
In this paper, we assume that an attacker is able to obtain the EHD between the return address and $key\_2$ through debugging and also can find available gadgets in the program. In what follows, we will evaluate the probability that the attacker launches an ROP attack.

Assume that the attacker gets the return address $R_i$ and the $EHD_i$ between $R_i$ and the $key\_2$, and can find a $gadget_j$ whose address $R_j$ has $x$-bit difference with $R_i$. If the attacker wants to jump to the $gadget_j$, he/she must first guess $m$ and then guess the $EHD_j$ between $R_j$ and $key\_2$. Since $m$ is set to be the last $k$-bit in the $key\_2$ in this paper, and $m$ ranges from 0 to $2^k-1$, the probability that $m$ is guessed correctly is ${(\frac{1}{2})}^k$.

If $x$ is odd, $HD_j$ has $x + 1$ possible values ($HD_i$ $\pm$ 1, $HD_i$ $\pm$ 3, $HD_i$ $\pm$ 5, ..., $HD_i$ $\pm$ $x$); If $x$ is even, $HD_j$ has $x+1$ possible values ($HD_i$ $\pm$ 0, $HD_i$ $\pm$ 2, $HD_i$ $\pm$ 4, ..., $HD_i$ $\pm$ $x$). Therefore, the probability that the attacker guesses the correct $HD_j$ between $R_j$ and $key\_2$ is $\frac{1}{x+1}$.

After correctly guessing $m$ and getting the correct $HD_i$, the attacker can guess the $EHD_j$. The probability that the attacker guesses the correct $EHD_j$ between $R_j$ and $key\_2$ would be ${(\frac{1}{2})}^k$ * $\frac{1}{x+1}$, and finally, the probability that the attacker could successfully perform an ROP attack is

\begin{equation}\label{(2)}
P = {[(\frac{1}{2})^k \ast (\frac{1}{x+1})]}^n
\end{equation}
where $n$ denotes the number of gadgets in an ROP gadgets chain.

Take an extreme case that $x$ is equal to 1 for example. The probability that attackers guess the correct $m$ is ${(\frac{1}{2})}^k$, and the probability that attackers guess the correct $HD_j$ between $R_j$ and $key\_2$ is $\frac{1}{2}$. Hence, the probability that attackers guess the correct $EHD_j$ between $R_j$ and $key\_2$ would be ${(\frac{1}{2})}^{k+1}$. In such worst and extreme case, the probability that an attacker could successfully perform an ROP attack is ${(\frac{1}{2})}^{n*(k+1)}$. It is difficult to launch an attack when $n*(k+1)$ is greater than 32.

\subsection{Advanced CRAs}
As an advanced CRA, the full-function reuse attack can utilize full functions as gadgets to implement attacks. Since there are many indirect call/jmp instructions existing in a program, attackers may use them to conduct the full-function CRA. For example, RIPE \cite{Wilander2011} contains 80 attacks that use indirect call instructions. If an attacker uses indirect call/jmp instructions to conduct thus advanced CRAs, most current defenses such as ASLR \cite{Bhatkar2005,Pappas2012,Wartell2012}, shadow stack \cite{Frantzen2001,Dang2015,Intel2017}, gadgets checking \cite{Chen2009} and CFI\cite{Qiu2017,Qiu2016,Das2016a} would be bypassed. For example, shadow stack techniques \cite{Das2016a} allow the return address to be any address in the shadow stack to avoid an exception thrown in the case of multithread and \emph{longjmp()}, so it is vulnerable to return-based full-function CRAs.

It is possible for attackers to bypass our ROP defense with the return-based full-function reuse. First, in the attack preparation phase, attackers traverse the program to find out all the available full-functions as gadgets. Then, they record these full-functions' return addresses and the corresponding EHDs when the program is running. Finally, attackers replace the return address and EHD in the stack with one of recorded available full-functions' return addresses and the corresponding EHD to hijack the control flow. This advanced CRAs can be prevented with our proposed key-updating method which can invalidate previous recorded addresses and EHDs.

\begin{figure}
\centerline{\includegraphics[width=\linewidth]{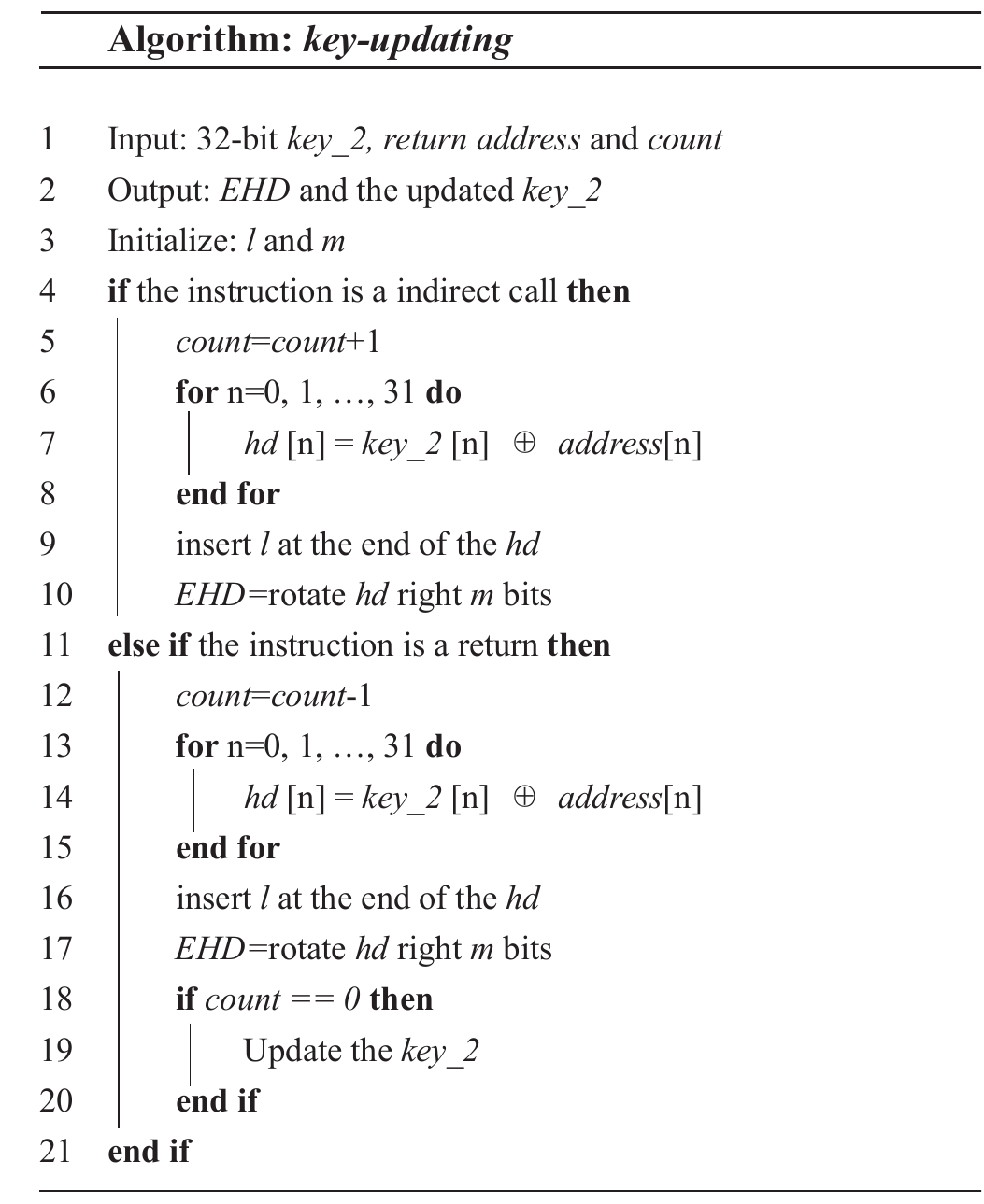}}
\label{fig9}
\end{figure}

The key-updating algorithm is shown above. The inputs to the algorithm are 32-bit PUF key $key\_2$, return address and the \emph{count} of counter. \emph{Count} is initialized to 0 when the key is generated. The outputs of the algorithm are the encrypted Hamming distance (See Section V.A) and the updated key. At the beginning of the algorithm, \emph{l} is initialized to the first $l$-bit of $key\_2$, and \emph{m} is initialized to the last $k$ bits of $key\_2$. A counter is used to record the number of times of the $key\_2$ used. When the call instruction is executed, the counter is increased by 1; when the ret instruction is executed, the counter is decreased by 1. When the counter is reduced to 0, the key will be updated immediately.

However, our defense mechanism is vulnerable to a more advanced CRA, COOP \cite{Schuster2015}. As a new emerged full-function reuse attack in C ++ applications, COOP uses virtual functions as gadgets and does not need to modify the function's return address so that HCIC is unable to detect such attack. We therefore assume auxiliary protections for virtual calls. For example, we assume that VTrust is deployed with about additional 0.72\% performance overhead \cite{zhang2016b}. In other words, we assume virtual function calls are well-protected. In addition, non-control data attacks \cite{chen2005}\cite{Carlini2015} tamper with or leak security sensitive memory, which is not directly used in control transfer instructions. Therefore, our approach, as well as all other control flow integrity methods, cannot prevent non-control data attacks. Usually, memory safety enforcement needs to be deployed to prevent the non-control-data attacks. However, the deployment of current memory safety proposals incurs high performance overhead. Development of practical memory safety defense is an active research area, which are out of the scope of this paper.

\subsection{Side channel attacks}
Side channel attack is to statistically analyze the electromagnetic emanation, power consumption or time of the cryptographic devices to gain knowledge about integrated secrets \cite{Zhang2014}. It is well-known that any key-based security mechanisms would be vulnerable to side channel attacks unless appropriate countermeasures are taken. In this paper, since the PUF key can be dynamically updated, our proposed HCIC mechanism would be less vulnerable to side channel attacks than traditional cryptographic keys which rely on permanent secure storage. However, our approach is not completely side channel attack free. In future, we plan to conduct the experiments on FPGAs to evaluate the resiliency of this technique to side channel attacks in detail.

\section{Experimental Results and Analysis}
The SPEC CPU2006 \cite{Henning2006}, BioBench  \cite{Albayraktaroglu2005}, MiBench \cite{Guthaus2001} and Stream benchmarks  \cite{McCalpin} are used in our experiment. These benchmarks are compiled using the GNU GCC version 4.9.2 at O3 optimization level on Ubuntu-15.04. Pin \cite{Naftaly2012} is used to get the target addresses of jmp and call instructions and the instructions at the target addresses for pre-processing of benchmarks. We use RIPE \cite{Wilander2011}, which contains 850 buffer overflow attacks, to evaluate the defense capability of our proposed mechanism.

\begin{table}[htbp]
  \newcommand{\tabincell}[2]{\begin{tabular}{@{}#1@{}}#2\end{tabular}}
  \caption{ROP Gadgets Reduction Tested by ROPGadget-v5.4}
  \label{Table1}
  \centering
  \begin{tabular}{|c|c|c|c|}
  \hline
  \tabincell{c}{\textbf{Benchmark}} & \tabincell{c}{\textbf{Total}\\\textbf{Gadgets}} & \tabincell{c}{\textbf{Allowed}\\\textbf{Gadgets}}&\tabincell{c}{\textbf{ROP Gadgets}\\\textbf{Elimination rate}} \\\hline
           mcf&      11927 &        0 &      100\%  \\ \hline
        hmmer &      15802 &        0 &      100\%  \\ \hline
   libquantum &      12531 &        0 &      100\%  \\ \hline
      h264ref &      17820 &        0 &      100\%  \\ \hline
          lbm &      12304 &        0 &      100\%  \\ \hline
     blowfish &      12140 &        0 &      100\%  \\ \hline
       phylip &      13466 &        0 &      100\%  \\ \hline
     specrand &      11591 &        0 &      100\%  \\ \hline
       stream &      11549 &        0 &      100\%  \\ \hline
    basicmath &      14664 &        0 &      100\%  \\ \hline
     patricia &      13188 &        0 &      100\%  \\ \hline
          sha &      13304 &        0 &      100\%  \\ \hline
   \textbf{Average}&            &          &      \textbf{100\% } \\ \hline
\end{tabular}
\end{table}

\subsection{Evaluation on RIPE Benchmark Attacks}

RIPE consists of 850 buffer-overflow attacks which can bypass ASLR and perform code injection, return-into-libc, and ROP on the stack, heap, BSS, and data segment. Our test results show that in the case of disabling DEP, 419 attacks out of 850 attacks in RIPE can be successful. Among the 419 attacks, 339 of them tamper the return addresses of ret instructions, such as code injection, return-into-libc, and ROP. Since our defense mechanism limits the return address to be the address of the next instruction of the corresponding call instruction by computing and matching EHD, these attacks get detected.

\subsection{ Gadgets Reduction}
We use the gadgets reduction as a metric to evaluate our defense mechanism. In general, attackers use the gadgets to perform CRAs, so the allowed gadgets reduction is one of the important metrics to evaluate a defense mechanism. For example, the average gadgets reduction for a previous work \cite{Das2016a} is 99.381\%. The reason of the allowed gadgets still existing in this defense mechanism is that there are some BBs can be exploited to perform an ROP. HCIC limits the return address to be the address of the next instruction of the corresponding call instruction, so the attacker is difficult to use BBs to perform ROP attacks. We use ROPGadget-v5.4 \cite{Salwan2011} to scan the binary to get all ROP gadgets in the program and get the number of allowed ROP gadgets which are used to bypass the defense mechanism and perform ROP attacks. Table I gives the number of ROP gadgets, allowed gadgets and the gadgets reduction rate for different benchmarks with HCIC. The test results show that HCIC can effectively reduce the allowed gadgets (the allowed gadgets are 0 and the gadgets reduction rate achieves 100\%).

\subsection{Average Indirect Target Reduction (AIR)}
In general, attackers hijack the normal control flow of the program and perform CRAs by tampering the target addresses of control flow instructions. Therefore, reducing the number of indirect targets can reduce the successful probability that attackers conduct CRAs. So, the reduction of indirect targets is one of the important metrics to evaluate a defense mechanism. The average indirect target reduction (AIR) metric\cite{Zhang2015a} is used to evaluate the reduction of indirect targets, as Eq. (3) shows.
\begin{equation}\label{2}
 AIR=1/n\sum_{j=1}^n(1-|T_j|/S)
\end{equation}
where, \emph{n} denotes the number of control flow instructions, $T_j$ denotes the number of indirect target addresses, \emph{S} denotes the size of binary code.

\begin{figure}
\centerline{\includegraphics[width=\linewidth]{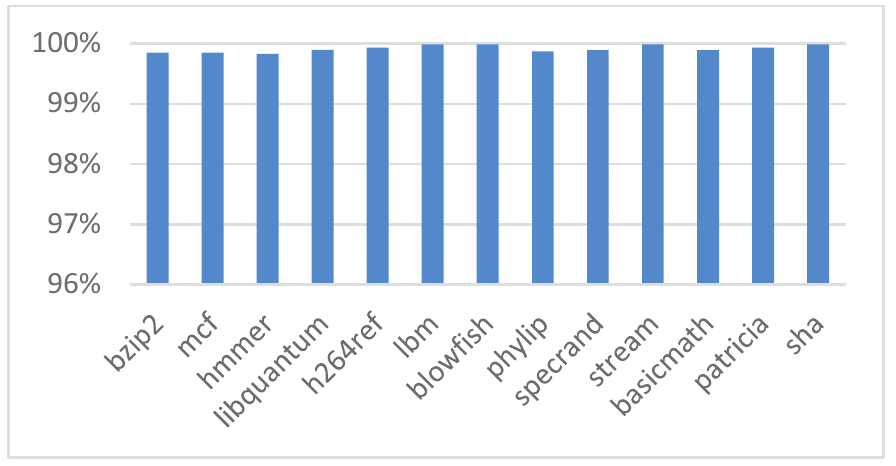}}
\caption{Average Indirect Target Reduction (AIR).}
\label{fig8}
\end{figure}

In HCIC, all call instructions target the beginning of functions, all jmp instructions target the starting address of BBs, and all ret instructions target the next instructions of the corresponding call instructions. Hence, the number of target addresses for call instructions is given by the number of functions, and the number of target addresses for jmp instructions is given by the number of BBs. For the ret instructions, the number of target addresses is always 1. Then Eq. (3) can be simplified to Eq. (4).
\begin{equation}\label{3}
  AIR=1-(|T_{call}|+|T_{jmp}|+1)/3S
\end{equation}
where, $T_{call}$ is the number of functions and the $T_{jmp}$ is the number of BBs. Fig. \ref{fig8} shows the estimation of AIR on benchmarks with HCIC. The reduction of indirect targets is greater than 99.8\%, which means that HCIC can eliminate almost all indirect targets.

\subsection{False positive/negative}
Typical CFI implementations require a precise CFG of the program. However, the generation of precise CFGs for real-word software remains an open research problem. Therefore, all CFI-based defenses that require the precise CFG may generate false positive. The CFG includes basic blocks (BB) information and the execution flow between BBs. Our approach only needs BBs information in CFG, which eliminates the need of source code analysis and static analysis. If the legal target addresses of some call or jmp instructions are not covered by the profiled CFG, HCIC would produce false positives when the program decrypts unencrypted instructions at those addresses. The false-positive can be reduced with a CFG as precise as possible. In order to get a CFG as precisely as possible, during profiling, to increase the coverage, we run each benchmark thousands of times with different set of valid inputs to get the possible target addresses of all the jmp and call instructions. On the other hand, if attackers can use full-functions as gadgets to launch advanced CRAs, the false-negative will occur. However, most of advanced CRAs can be detected with HCIC. Therefore, in theory, the false negative rate would be extremely low.

\begin{table}[htbp]
  \newcommand{\tabincell}[2]{\begin{tabular}{@{}#1@{}}#2\end{tabular}}
  \caption{ Runtime overhead}
  \label{Table2}
  \centering
  \begin{tabular}{|c|c|c|c|}
  \hline
  \tabincell{c}{\textbf{Benchmark}} & \tabincell{c}{\textbf{Original}\\\textbf{runtime(s)}} & \tabincell{c}{\textbf{Result}\\\textbf{runtime(s)}}&\tabincell{c}{\textbf{Runtime}\\\textbf{overhead}} \\\hline
           mcf&       2.652508 &        2.684364 &       1.2\%  \\ \hline
        hmmer &       0.003413 &        0.003450 &      1.08\%  \\ \hline
   libquantum &       0.004010 &        0.004041 &      0.77\%  \\ \hline
      h264ref &      29.620388 &       30.010754 &      1.31\%  \\ \hline
          lbm &       2.753622 &        2.763091 &      0.34\%  \\ \hline
     blowfish &       0.000741 &        0.000752 &       1.5\%  \\ \hline
       phylip &       0.002136 &        0.002159 &       1.1\%  \\ \hline
     specrand &       0.023326 &        0.023489 &       0.7\%  \\ \hline
       stream &       1.105379 &        1.111355 &      0.54\%  \\ \hline
    basicmath &       0.284601 &        0.284694 &      0.03\%  \\ \hline
     patricia &       0.076869 &        0.077837 &      1.25\%  \\ \hline
          sha &       0.037029 &        0.037612 &      1.57\%  \\ \hline
   \textbf{Average}&           &                 &      \textbf{0.95\% } \\ \hline
\end{tabular}
\end{table}

\begin{table}[htbp]
  \newcommand{\tabincell}[2]{\begin{tabular}{@{}#1@{}}#2\end{tabular}}
  \caption{ Binary size overhead}
  \label{Table3}
  \centering
  \begin{tabular}{|c|c|c|c|}
  \hline
  \tabincell{c}{\textbf{Benchmark}} & \tabincell{c}{\textbf{Original}\\\textbf{Size(B)}} & \tabincell{c}{\textbf{Result}\\\textbf{Size(B)}}&\tabincell{c}{\textbf{Binary Size}\\\textbf{overhead}} \\\hline
           mcf&        761244 &        767145 &       0.7753\%  \\ \hline
        hmmer &       1119396 &       1130997 &       1.0364\%  \\ \hline
   libquantum &        853244 &        859221 &       0.7005\%  \\ \hline
      h264ref &        780604 &        781334 &       0.0935\%  \\ \hline
          lbm &        781872 &        786030 &       0.5318\%  \\ \hline
     blowfish &        764524 &        767004 &       0.3244\%  \\ \hline
       phylip &        858452 &        864139 &       0.6625\%  \\ \hline
     specrand &        743580 &        748150 &       0.6147\%  \\ \hline
       stream &        752128 &        758546 &       0.8534\%  \\ \hline
    basicmath &        803960 &        812411 &       1.0512\%  \\ \hline
     patricia &        748032 &        756532 &       1.1363\%  \\ \hline
          sha &        747808 &        759476 &       1.5603\%  \\ \hline
   \textbf{Average}&           &                 &      \textbf{0.7783\% } \\ \hline
\end{tabular}
\end{table}

Several tools have been developed to extract the BB information from a program such as Pin and Valgrind. In our experiments, we first use Pin to collect all destination addresses of call and jmp instructions before the program is loaded into memory. Then, we encode the first instruction with the XOR operation at the target address of call and jmp instructions to prevent JOP. The experimental results show that the false positive is 0\% with HCIC. Besides, we evaluated the false negative by analyzing gadgets and indirect targets reduction. Our experimental results show that HCIC can reduce the ROP gadgets and indirect targets by 100\% and 99.8\%, respectively.

\subsection{Performance and Binary Size Overhead}
 HCIC computes and matches the EHD, and decrypts the instructions at the target addresses of the call and jmp instructions when the program is running, which would generate the runtime overhead. In our experiments, we insert an ``andl \%eax, \%eax'' instruction in the beginning and ending of each function, and insert an ``andl \%eax, \%eax'' instruction before each call and jmp instruction. The reason for inserting the ``andl \%eax, \%eax'' instruction is that this instruction will not modify the value of the register and the program can be executed normally. As shown in Table II, the average runtime overhead is 0.95\%, which is far less than the performance overhead of traditional CFIs.

Because the sequential execution of encrypted instructions will produce the false positive (i.e., the first loop of the do-while statement), we insert a jmp instruction before the instruction at the destination address of a jmp instruction to ensure that the program can run normally. The destination address of the inserted jmp instruction is the next encrypted instruction's address. We just insert one instruction before the instruction at the destination address of the jmp instruction. Therefore, HCIC produces very low binary size overhead. As shown in Table III, the average binary size overhead is just 0.78\%.

\begin{table*}[htbp]
  \centering
  \begin{threeparttable}
  \newcommand{\tabincell}[2]{\begin{tabular}{@{}#1@{}}#2\end{tabular}}
  \caption{Comparison of Security and Practicality}
  \label{Table4}
  \begin{tabular}{|c|c|c|c|c|c|c|}
  \hline
  \multirow{2}{*}{}&\multicolumn{2}{c|}{Security} &\multicolumn{4}{c|}{Practicality} \\\cline{2-7}
   &\tabincell{c}{Level}&\tabincell{c}{{Key}\\{leakage}}&\tabincell{c}{{ISA}\\{Extensions}}&\tabincell{c}{{Compiler}\\{modification}}
   &\tabincell{c}{{Binary Size}\\{increasement}}&\tabincell{c}{{Performance}\\{Overhead}} \\ \hline
   \tabincell{c}{{CFI}\\{(CCS'05 \cite{Abadi2005})}}&      I & $\backslash$  &  N &  N &  High(8\%)  &  High(21\%)   \\ \hline
   \tabincell{c}{{CCFI}\\{(CCS'15 \cite{Mashtizadeh2015})}}&     I &           N  &  Y &  Y &  Y\tnote{1}  &  High(52\%)   \\ \hline
   \tabincell{c}{{SOFIA}\\{(DATE'16 \cite{Sullivan2016})}}&     II &           N  &  N &  N &  Y\tnote{1}  &  High(110\%)   \\ \hline
   \tabincell{c}{{LEA}\\{(DAC16' \cite{Qiu2016})}}&   II &           Y  &  N &  N &  Low(0.78\%)  &  Low(0.9\%)   \\ \hline
   \tabincell{c}{{HECFI}\\{(DAC'16 \cite{Sullivan2016})}}& III & $\backslash$  &  Y &  Y &High(13.5\%)&  Low(1.75\%)   \\ \hline
   \tabincell{c}{{LEA-AES}\\{(TCAD'17 \cite{Qiu2017})}}&     II &           N  &  Y &  Y &  Low(0.78\%)  & Low(3.2\%)   \\ \hline

  \tabincell{c}{\textbf{Our proposed}\\\textbf{HCIC}}&\textbf{II} &\textbf{N} &\textbf{N} &\textbf{N}&\textbf{Low(0.78\%)} &Low(\textbf{0.95}\%)   \\ \hline
\end{tabular}

\begin{tablenotes}
        \footnotesize
        \item[1] The authors did not give the data of binary size overhead.
        \end{tablenotes}
\end{threeparttable}
\end{table*}

\subsection{Comparison of Security and Practicality}
We compare HCIC with recent methods \cite{Abadi2005}\cite{Mashtizadeh2015}\cite{Clercq2016}\cite{Qiu2016}\cite{Sullivan2016}\cite{Qiu2017}\cite{zhang2016b}. Security and practicality are two most important metrics to evaluate current CRA defenses. Practicality is evaluated by ISA extensions, compiler modification, binary size increasement and performance overhead. Security is evaluated by the key leakage and security level.
We divide the security level of defense mechanisms into the following four levels:
\begin{itemize}
  \item Level-I: Only defend against ROP;
  \item Level-II: defend against ROP and JOP;
  \item Level-III: defend against ROP, JOP and some advanced CRAs such as COOP;
  \item Level-IV: defend against all potential CRAs.
\end{itemize}

As shown in Table IV, HCIC incurs low performance and binary overheads, and does not need to extend ISAs and modify compilers. Besides, HCIC can achieve the level-II without leaking the key. It is worth noting that HCIC can be enriched with VTrust \cite{zhang2016b} (VTrust protects virtual call only and is unable to prevent against ROP and JOP) to defend against COOP with additional 0.72\% performance overhead. Obviously, our proposed HCIC shows the best balance between security and practicality.

\section{Conclusion}
This paper proposes a hardware-assisted CFI checking technique to resolve the vulnerabilities that current software-based CFI incurs high performance overhead and hardware-based may require extending the existing processors' ISAs or suffer some security vulnerabilities. The key technique involves two control flow verification mechanisms. The first one is to compute EHDs between the PUF response and the return addresses, then verifies whether the EHDs are matched to make attackers impossible to return between gadgets, thus resisting ROP attacks. The second one is to perform the linear XOR operation between the PUF key and the instructions at target addresses of call and jmp instructions once the executable binary is loaded into memory, then the runtime linear decryption operation can be done when jmp and call instructions are executed, thus defeating JOP attacks. The experiment results show that the proposed new technique incurs extremely low performance overhead (average 0.95\%) and binary size overhead (average 0.78\%), which are much lower than traditional CFI approaches. Exception analysis shows that our proposed defense does not produce anomalies when some exceptional cases occurred. Security analysis also shows that the proposed method is sound and secure against code reuse attacks with zero false positive and negative rates.

Coarse-grained CFI is more vulnerable to function reuse attacks than fine-grained CFI which validates call sites beyond just the target address considering elements such as expected VTable type, validating number of arguments, or even argument types, for a given indirect call. A key tradeoff between the two approaches is performance, as introducing too much complexity into a CFI policy can add significant overhead. Considering this, current deployed CFI schemes in industry are almost coarse-grained CFI such as Microsoft's Control Flow Guard and Intel's Control-flow Enforcement Technology. HCIC is able to prevent ROP, JOP and return-based full-function reuse attacks, and can be enriched with auxiliary protections to prevent virtual function reuse with additional small performance penalty. Our proposed method shows a good balance between security and practicality.

\begin{IEEEbiography}[{\includegraphics[width=1in,height=1.25in,clip,keepaspectratio]{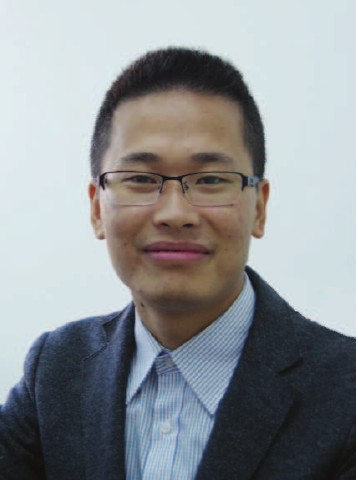}}]{Jiliang Zhang} received the Ph.D. degree in Computer Science and Technology from Hunan University, Changsha, China in 2015. From 2013 to 2014, he worked as a Research Scholar at the Maryland Embedded Systems and Hardware Security Lab, University of Maryland, College Park. From 2015 to 2017, he was an Associate Professor with Northeastern University, China. Since 2017, he has joined Hunan University. He has authored over 30 papers in refereed international conferences and journals such as the IEEE TRANSACTIONS ON INFORMATION FORENSICS AND SECURITY, IEEE TRANSACTIONS ON COMPUTER-AIDED DESIGN of Integrated Circuits and Systems, IEEE TRANSACTIONS ON VERY LARGE SCALE INTEGRATION, ACM Transactions on Design Automation of Electronic Systems, and ACM/IEEE Design Automation Conference. His current research interests include hardware/hardware-assisted security, artificial intelligence security and emerging technologies.
\end{IEEEbiography}


\begin{IEEEbiographynophoto}{Binhang Qi} received the B.S. degree in Information Security from Northeastern University, China in 2018. He is currently a visiting student at Hunan University, China. His research interests include hardware-assisted security.
\end{IEEEbiographynophoto}

\begin{IEEEbiographynophoto}{Zheng Qin}
received the Ph.D. degree in computer science from Chongqing University, Chongqing, China, in 2001. From 2010 to 2011, he served as a Visiting Scholar with the Department of Computer Science, Michigan State University. He is currently a Professor with the College of Computer Science and Electronic Engineering, Hunan University. He has authored over 80 papers in well-known journals and international conferences. His current research interests include network and data security, data analytics and applications, machine learning, and applied cryptography.
\end{IEEEbiographynophoto}

\begin{IEEEbiographynophoto}{Gang Qu}
received the B.S. and M.S. degrees in mathematics from the University of Science and Technology of China, in 1992 and 1994, respectively, and the Ph.D. degree in computer science from the University of California, Los Angeles, in 2000. Upon graduation, he joined the University of Maryland at College Park, where he is currently a professor in the Department of Electrical and Computer Engineering and Institute for Systems Research. His primary research interests are in the area of embedded systems and VLSI CAD with focus on low power system design and hardware related security and trust. He studies optimization and combinatorial problems and applies his theoretical discovery to applications in VLSI CAD, wireless sensor network, bioinformatics, and cybersecurity.
\end{IEEEbiographynophoto}



\vfill

\end{document}